\documentclass{elsarticle}

\usepackage{lineno,hyperref}

\usepackage{amssymb}
\usepackage{latexsym}
\usepackage{amsmath}
\usepackage{graphicx}
\usepackage{pifont}
\usepackage{natbib}
\usepackage{geometry}
\usepackage{fleqn}

\begin{document}
\begin{frontmatter}


\title{Variational Methods For Phononic Calculations}

\author[mymainaddress]{Yan Lu\corref{mycorrespondingauthor}}
\cortext[mycorrespondingauthor]{Corresponding author}
\ead{ylu50@hawk.iit.edu}
\author[mymainaddress]{Ankit Srivastava}

\address[mymainaddress]{Department of Mechanical, Materials, and Aerospace Engineering, Illinois Institute of Technology, Chicago, IL, 60616
USA}

\begin{abstract}
Three fundamental variational principles used for solving elastodynamic eigenvalue problems are studied within the context of elastic wave propagation in periodic composites (phononics). We study the convergence of the eigenvalue problems resulting from the displacement Rayleigh quotient, the stress Rayleigh quotient and the mixed quotient. The convergence rates of the three quotients are found to be related to the continuity and differentiability of the density and compliance variation over the unit cell.  In general, the mixed quotient converges faster than both the displacement Rayleigh and the stress Rayleigh quotients, however, there exist special cases where either the displacement Rayleigh or the stress Rayleigh quotient shows the exact same convergence as the mixed-method. We show that all methods converge faster for smoother material property variations,  but when density variation is rough, the difference between the mixed quotient and stress Rayleigh quotient is higher and similarly, when compliance variation is rough, the difference between the mixed quotient and displacement Rayleigh quotient is higher. Since eigenvalue problems such as those considered in this paper tend to be highly computationally intensive, it is expected that these results will lead to fast and efficient algorithms in the areas of phononics and photonics.
\end{abstract}

\end{frontmatter}

\section{Introduction}
The periodic modulation of stress wave existing in periodic composites results in exotic dynamic response. The phononic band-structure\cite{martinezsala1995sound} induced by the periodic modulation of stress waves has close similarities with areas like electronic band theory\cite{bloch1928quantum} and photonics\cite{ho1990existence}. The rich wave-physics resulting from the periodic modulations offers potential for novel applications such as refractive acoustic devices\cite{cervera2001refractive}, ultrasound tunneling\cite{yang2002ultrasound}, waveguiding\cite{khelif2003trapping}, reversed Doppler effect\cite{reed2003reversed}, sound focusing\cite{yang2004focusing}, hypersonic control\cite{gorishnyy2005hypersonic},  negative refraction\cite{sukhovich2008negative}, gradient-index refraction\cite{lin2009gradient}, etc. The first step in realizing these applications is to calculate the phononic band-structure. Additionally, some research areas such as phononic band-structure optimization\cite{sigmund2003systematic,rupp2007design,bilal2011ultrawide,diaz2005design,halkjaer2006maximizing} and inverse problems in dynamic homogenization\cite{nemat2011homogenization} depend heavily on the speed, efficiency, accuracy and versatility of the band-structure calculating algorithm. There exist several techniques by which band-structures of photonic and phononic composites can be calculated. The plane wave expansion method (PWE)\cite{ho1990existence,leung1990full,zhang1990electromagnetic} is easy to implement but converges slowly when material properties show large contrast. The multiple scattering method\cite{kafesaki1999multiple,mei2005multiple} can precisely predict the band structure, but is subject to geometry limitations. The finite element method (FEM)\cite{white1989finite,veres2012complexity,hladky1991analysis,hussein2006mode} is derived from variational principles and it is widely used for computing phononic bandstructures of various geometries. Hussein and Hulbert\cite{hussein2006mode} have presented a mixed finite element approach based on the variation of displacement and strain field, which shares the same mathematical background (Hu-Washizu variational theorem) as the methods presented in this paper. Some other methods are the finite difference time domain method (FDTD)\cite{chan1995order,tanaka2000band} and secondary expansions such as the reduced Bloch mode expansion\cite{hussein2009reduced} method, etc.

Convergence rates of three variational principles, the displacement Rayleigh quotient, where the displacement field is varied, the stress Rayleigh quotient, where the stress field is varied, and the mixed quotient\cite{nemat1972general,nemat1975harmonic,minagawa1976harmonic,nemat2011homogenization}, where both the displacement and stress fields are varied, are considered in this paper. In addition, we also compare the convergence rates of the variational methods with the most popular band structure algorithms (PWE and displacement based FEM). The mixed quotient was proposed by Nemat-Nasser\cite{nemat1972general} in 1972, which was derived from the works of Hellinger\cite{hellinger1914allgemeinen}, Prange\cite{prange1916unpublished}, Reissner\cite{reissner1950elasticity,reissner1953finite}, Hu\cite{hu1955variational}, Washizu\cite{washizu1955variational}. Nemat-Nasser\cite{nemat1975harmonic} showed that when compliance is constant, the mixed quotient reduces to the displacement Rayleigh quotient and when density is constant, the mixed quotient reduces to the stress Rayleigh quotient. Babuska and Osborn\cite{babuska1978numerical} related the convergence behavior of the three variational principles to the smoothness of the compliance and density functions and showed that the mixed quotient, in general, converges faster than the other two methods. Due to its excellent accuracy and efficiency, the mixed variation principle has been recently applied to the study of wave refraction in periodic elastic composites\cite{nemat2015refraction,nemat2015anti}.

In this paper, the three variational principles which solve the phononic eigenvalue problems are presented in Section \ref{stateproblem}. The algorithms derived from the variational principles are suitable for applications to arbitrary unit cells. We present the detailed formulation of the eigenvalue problems and their convergence behavior under different compliance and density distribution for 1-D and 2-D periodic composites. In our calculation we consider the effects of different function continuity and differentiability conditions of compliance and density on the convergence rates of the three formulations.

\section{Statement of the problem}\label{stateproblem}
Phononic computations seek to evaluate the essential properties of sound/stress waves traveling in periodic structures. These properties include the phononic band-structure evaluated along the boundary of the Irreducible Brillouin zone (IBZ) of 1-, 2-, and 3-D composites, their equi-frequency contours, and the associated density of states. However, all these properties emerge from the solution of the fundamental eigenvalue problem which is associated with the elastodynamics of periodic structures. In the following subsections we define the essential properties of the periodic domain under consideration, eigenvalue problem associated with wave propagation in this periodic domain, and the variational methods which can be employed for its solution.

\subsection{Periodic Domain}
\begin{figure}[htp]
\centering
\includegraphics[scale=1]{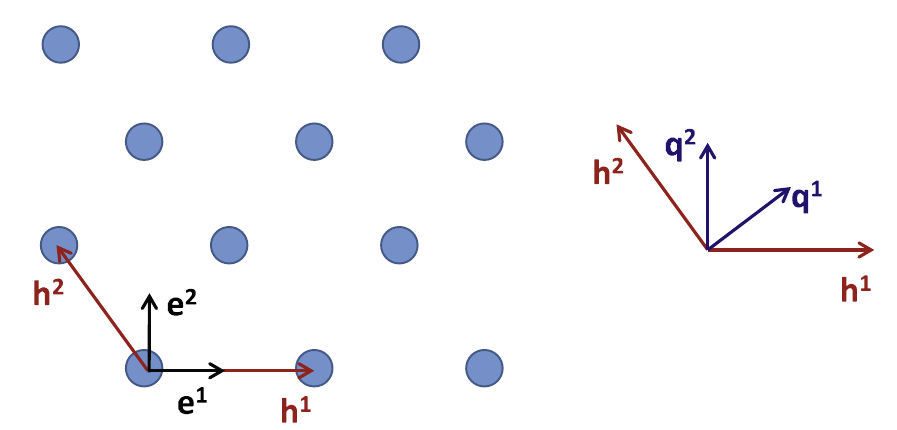}
\caption{Schematic of a 2-dimensional periodic composite. The unit cell vectors ($\mathbf{h}^1,\mathbf{h}^2$), reciprocal basis vectors ($\mathbf{q}^1,\mathbf{q}^2$), and the orthogonal vectors ($\mathbf{e}^1,\mathbf{e}^2$) are shown.}\label{fVectors}
\end{figure}

In the following treatment repeated Latin indices mean summation, whereas, repeated Greek indices do not. Consider a general 3-dimensional periodic composite. The unit cell of the periodic composite is denoted by $\Omega$ and is characterized by 3 base vectors $\mathbf{h}^i$, $i=1,2,3$. Any point within the unit cell can be uniquely specified by the vector $\mathbf{x}=H_i\mathbf{h}^i$ where $0\leq H_i\leq 1,\forall i$. The same point can also be specified in the orthogonal basis as $\mathbf{x}=x_i\mathbf{e}^i$. The reciprocal base vectors of the unit cell are given by:
\begin{equation}
\mathbf{q}^1=2\pi\frac{\mathbf{h}^2\times\mathbf{h}^3}{\mathbf{h}^1\cdot(\mathbf{h}^2\times\mathbf{h}^3)};\quad \mathbf{q}^2=2\pi\frac{\mathbf{h}^3\times\mathbf{h}^1}{\mathbf{h}^2\cdot(\mathbf{h}^3\times\mathbf{h}^1)};\quad \mathbf{q}^3=2\pi\frac{\mathbf{h}^1\times\mathbf{h}^2}{\mathbf{h}^3\cdot(\mathbf{h}^1\times\mathbf{h}^2)}
\end{equation}
such that $\mathbf{q}^i\cdot\mathbf{h}^j=2\pi\delta_{ij}$. Fig. (\ref{fVectors}) shows the schematic of a 2-D unit cell, indicating the unit cell basis vectors, the reciprocal basis vectors and the orthogonal basis vectors. The composite is characterized by a spatially varying stiffness tensor, $C_{jkmn}(\mathbf{x})$, and density, $\rho(\mathbf{x})$, which satisfy the following periodicity conditions:
\begin{equation}
C_{jkmn}(\mathbf{x}+n_i\mathbf{h}^i)=C_{jkmn}(\mathbf{x});\quad \rho(\mathbf{x}+n_i\mathbf{h}^i)=\rho(\mathbf{x})
\end{equation}
where $n_i$ $(i=1,2,3)$ are integers.
For wave propagation in such a periodic composite the wave vector is given as $\mathbf{k}=Q_i\mathbf{q}^i$ where $0\leq Q_i\leq 1,\forall i$. 

\subsection{Field Equations and Boundary Conditions}
For harmonic elastodynamic problems the equation of motion at any point $\mathbf{x}$ in $\Omega$ is given by
\begin{equation}\label{equationofmotion}
\sigma_{jk,k}=-\lambda\rho u_j
\end{equation}
where $\lambda=\omega^2$, and $\boldsymbol{\sigma}\exp\left[-i\omega t\right],\mathbf{u}\exp\left[-i\omega t\right]$ are the space and time dependent stress tensor and displacement vector respectively. The stress tensor is related to the strain tensor through the elasticity tensor, $\sigma_{jk}=C_{jkmn}\varepsilon_{mn}$, and the strain tensor is related to the displacement vector through the kinematic relation, $\varepsilon_{jk}=.5(u_{j,k}+u_{k,j})\equiv u_{(j,k)}$. Due to the periodicity of the composite the traction and displacement at any point $\mathbf{x}$ are related to the corresponding traction and displacement at another point, separated from the first by a unit cell, through Bloch relations. These relations serve as the homogeneous boundary conditions on $\partial\Omega$. If the Bloch wave vector is $\mathbf{k}$ then these boundary conditions are given by:
\begin{equation}\label{boundaryconditions}
u_j(\mathbf{x}+\mathbf{h}^i)=u_j(\mathbf{x})\exp\left[i\mathbf{k}\cdot\mathbf{h}^i\right];\quad t_j(\mathbf{x}+\mathbf{h}^i)=-t_j(\mathbf{x})\exp\left[i\mathbf{k}\cdot\mathbf{h}^i\right], \quad \mathbf{x}\in\partial\Omega
\end{equation}
where $t_j=\sigma_{jk}\nu_k$ are the components of the traction vector and $\boldsymbol{\nu}$ is the exterior normal vector on $\partial\Omega$. The $\mathbf{k},\lambda$ (or $\mathbf{k},\omega$) pairs which satisfy the equation of motion along with the Bloch boundary conditions constitute the solution to the phononic eigenvalue problem.

\subsection{Variational Methods for Approximate Solution}

The most straightforward way of finding a variational solution to the above problem is through the Rayleigh quotient method. This constitutes expressing the equation of motion in terms of displacement through the use of the constitutive and kinematic relations:
\begin{equation}\label{equationofmotionDisp}
\left[C_{jkmn}u_{m,n}\right]_{,k}=-\lambda\rho u_j
\end{equation}
It can be shown that the displacement field which satisfies the above equation of motion along with the Bloch boundary conditions minimizes the following functional (displacement Rayleigh quotient):
\begin{equation}\label{dispR}
\lambda_u=\frac{\langle C_{jkmn}u_{m,n},u_{j,k}\rangle}{\langle\rho u_j,u_j\rangle}.
\end{equation}
The inner product is given by:
\begin{equation}
\langle u,v\rangle=\int_\Omega uv^*d\Omega
\end{equation}
where $v^*$ is the complex conjugate of $v$. The solutions to the phononic eigenvalue problem are calculated by finding the infimum of $\lambda_u$. 
An analogous single variable quotient can be derived by expressing the governing equation in terms of the stress field:
\begin{equation}\label{equationofmotionStr}
\left[\bar{\rho}\sigma_{jn,n}\right]_{,k}+\lambda D_{jkmn}\sigma_{mn}=0
\end{equation}
where $\bar{\rho}=1/\rho$ and $\mathbf{D}=\mathbf{C}^{-1}$. The above form leads the following functional (stress Rayleigh quotient):
\begin{equation}\label{sigmaR}
\lambda_\sigma=\frac{\langle \bar{\rho}\sigma_{jn,n},\sigma_{jk,k}\rangle}{\langle D_{jkmn}\sigma_{jk},\sigma_{mn}\rangle}
\end{equation}
which must be minimized to find the eigenvalue solutions of the phononic problem. Yet another variational formulation can be derived by considering a modified form of the governing equation:
\begin{eqnarray}\label{equationofmotionMix}
\nonumber u_{(j,k)}=D_{jkmn}\sigma_{mn}\\
\sigma_{mn,n}=-\lambda \rho u_m
\end{eqnarray}
The above form admits variations on both the displacement and the stress fields and leads to faster convergence of the eigenvalue solution.
The solution to (\ref{equationofmotionMix}) that satisfies the Bloch boundary conditions renders the following functional stationary (mixed quotient):
\begin{equation}\label{mixedvariational}
\lambda_{u\sigma}=\frac{\langle\sigma_{mn},u_{m,n}\rangle+\langle u_{j,k},\sigma_{jk}\rangle-\langle D_{jkmn}\sigma_{mn},\sigma_{jk}\rangle}{\langle\rho u_m,u_m\rangle}
\end{equation}

The three variational methods above can be derived from Hu-Washizu variational theorem, which includes all the equations for the linear theory of elasticity (without inertial forces). The detailed derivation is presented in \cite{nemat1972general} and the quotient formulations have been explicitly given in \cite{nemat1975harmonic}.
\subsection{Eigenvalue Forms}

The minimization problems mentioned in the previous section can be tackled by expanding the displacement and stress fields using chosen test functions to satisfy the boundary conditions and continuity conditions. :
\begin{equation}\label{approximation}
\bar{u}_j=\sum_{\alpha,\beta,\gamma}U^{\alpha\beta\gamma}_jf^{\alpha\beta\gamma}(\mathbf{x}),\quad \bar{\sigma}_{jk}=\sum_{\alpha,\beta,\gamma}S^{\alpha\beta\gamma}_{jk}f^{\alpha\beta\gamma}(\mathbf{x})
\end{equation}
The test functions are appropriately chosen to be orthogonal in the sense that $\langle f^{\alpha\beta\gamma},f^{\theta\eta\xi}\rangle$ is proportional to $\delta_{\alpha\theta}\delta_{\beta\eta}\delta_{\gamma\xi}$, $\boldsymbol{\delta}$ being the Kronecker delta. Substituting from (\ref{approximation})$^1$ to (\ref{dispR}) and setting the derivative of $\lambda_u$ with respect to the unknown coefficients, $U^{\alpha\beta\gamma}_j$, equal to zero, we arrive at the following system of linear homogeneous equations:
\begin{eqnarray}\label{equationshomogeneousDispR}
\langle\left[C_{jkmn}\bar{u}_{m,n}\right]_{,k}+\lambda_u\rho\bar{u}_j,f^{\theta\eta\xi}\rangle=0\nonumber\\
j,k,m,n=1,2,3
\end{eqnarray}
For the general 3-dimensional case, if trigonometric expansion terms are used and $\alpha,\beta,\gamma,\theta,\eta,\xi$ vary from $-M$ to $M$ then (\ref{equationshomogeneousDispR}) represents $3(2M+1)^3$ linear homogeneous equations in the $3(2M+1)^3$ unknown displacement coefficients. This eigenvalue problem can be cast into the following generalized matrix form:
\begin{equation}\label{eigenvalueMatrixdispR}
\mathbf{C}\mathbf{U}=-\lambda_u\boldsymbol{\rho}\mathbf{U}
\end{equation}
where $\mathbf{C}$ and $\boldsymbol{\rho}$ are $3(2M+1)^3\times 3(2M+1)^3$ matrices and $\mathbf{U}$ is the eigenvector consisting of the displacement coefficients used in the expansion. Similar sets of eigenvalue equations can be derived for the stress Rayleigh quotient and the mixed quotient. For the stress Rayleigh quotient the equations are:
\begin{eqnarray}\label{equationshomogeneousStressR}
\langle\left[\bar{\rho}\bar{\sigma}_{jn,n}\right]_{,k}+\lambda_\sigma D_{jkmn}\bar{\sigma}_{mn},f^{\theta\eta\xi}\rangle=0\nonumber\\
j,k,m,n=1,2,3
\end{eqnarray}
which have been derived by substituting from (\ref{approximation})$^2$ to (\ref{sigmaR}) and setting the derivative of $\lambda_\sigma$ with respect to the unknown coefficients, $S^{\alpha\beta\gamma}_{jk}$, equal to zero. Taking into account the symmetry of the stress tensor the above represent $6(2M+1)^3$ linear homogeneous equations in the $6(2M+1)^3$ unknown stress coefficients, when trigonometric expansion terms are used. This eigenvalue problem can be cast into the following matrix form:
\begin{equation}\label{eigenvalueMatrixstressR}
\bar{\boldsymbol{\rho}}\mathbf{S}=-\lambda_\sigma\mathbf{D}\mathbf{S}
\end{equation}
where $\mathbf{D}$ and $\bar{\boldsymbol{\rho}}$ are $6(2M+1)^3\times 6(2M+1)^3$ matrices and $\mathbf{S}$ is the eigenvector consisting of the stress coefficients used in the expansion. It is clear, therefore, that the matrices characterizing the stress Rayleigh eigenvalue problem are 4 times bigger than the ones characterizing the displacement Rayleigh quotient. The eigenvalue problem resulting from the mixed-quotient can be derived by substituting from (\ref{approximation}) to (\ref{mixedvariational}) and setting the derivative of $\lambda_{u\sigma}$ with respect to the unknown displacement and stress coefficients equal to zero. Taking into account the symmetry of the stress tensor we arrive at the following $9(2M+1)^3$ linear homogeneous equations in the $9(2M+1)^3$ unknown displacement and stress coefficients, when trigonometric expansion terms are used:
\begin{eqnarray}\label{equationshomogeneous}
\langle\bar{\sigma}_{mn,n}+\lambda_{u\sigma}\rho\bar{u}_m,f^{\theta\eta\xi}\rangle=0\nonumber\\
\langle D_{jkmn}\bar{\sigma}_{mn}-\bar{u}_{(j,k)},f^{\theta\eta\xi}\rangle=0\nonumber\\
j,k,m,n=1,2,3
\end{eqnarray}
These equations can be written in the following equivalent matrix form:
\begin{eqnarray}\label{equationshomogeneousMatrix}
\nonumber \mathbf{HS}+\lambda_{u\sigma}\mathbf{\Omega U}=0\\
\mathbf{\Phi S}+\mathbf{H^*U}=0
\end{eqnarray}
Matrices $\mathbf{H},\mathbf{\Omega},\mathbf{\Phi},\mathbf{H}^*$ contain the integrals of the various functions appearing in (\ref{equationshomogeneous}). The above system of equations can be recast into the following form:
\begin{equation}\label{eigenvalueproblem}
\mathbf{H}\mathbf{\Phi}^{-1}\mathbf{H}^*\mathbf{U}=\lambda_{u\sigma}\mathbf{\Omega}\mathbf{U}
\end{equation}
\subsection{Trigonometric Expansion}

To approximate the stress and displacement fields in (\ref{approximation}), we use test functions of the following form:
\begin{equation}\label{testfunction0}
f^{\alpha\beta\gamma}(\mathbf{x})=\exp\left[i(\mathbf{k}\cdot\mathbf{x}+2\pi[\alpha H_1+\beta H_2+\gamma H_3])\right]
\end{equation}
where $\mathbf{x}=H_j\mathbf{h}^j$. One way for the displacement and stress field to satisfy the boundary conditions, (\ref{boundaryconditions}), is to choose the test functions which satisfy the boundary conditions themselves. To show that these test functions satisfy the boundary conditions we note that $\mathbf{x}'=\mathbf{x}+\mathbf{h}^k=(H_j+\delta_{jk})\mathbf{h}^j$ and, therefore:
\begin{eqnarray}
f^{\alpha\beta\gamma}(\mathbf{x}')=\exp\left[i(\mathbf{k}\cdot\mathbf{x}+2\pi[\alpha H_1+\beta H_2+\gamma H_3])\right]\exp\left[i(\mathbf{k}\cdot(\mathbf{h}^j\delta_{jk}))\right]\exp\left[i(2\pi[\alpha\delta_{1k}+\beta\delta_{2k}+\gamma\delta_{3k}])\right]
\end{eqnarray}
Since $\alpha,\beta,\gamma$ are integers the last term equals unity and we have
\begin{equation}
f^{\alpha\beta\gamma}(\mathbf{x}+\mathbf{h}^k)=f^{\alpha\beta\gamma}(\mathbf{x})\exp\left[i(\mathbf{k}\cdot\mathbf{h}^k)\right]
\end{equation}
showing that test functions satisfy the boundary conditions. To show orthogonality we note that
\begin{eqnarray}
\nonumber \langle f^{\alpha\beta\gamma},f^{\theta\eta\xi}\rangle=\int_\Omega \exp\left[i(\mathbf{k}\cdot\mathbf{x}+2\pi[\alpha H_1+\beta H_2+\gamma H_3])\right] \exp\left[-i(\mathbf{k}\cdot\mathbf{x}+2\pi[\theta H_1+\eta H_2+\xi H_3])\right]d\Omega\\
=\int_\Omega \exp\left[i(2\pi[(\alpha-\theta) H_1+(\beta-\eta) H_2+(\gamma-\xi) H_3])\right]d\Omega\propto\delta_{\alpha\theta}\delta_{\beta\eta}\delta_{\gamma\xi}
\end{eqnarray}
showing the orthogonality of the test functions.

\section{Study of Convergence Rates}\label{convergentrate}

We are primarily interested in studying the convergence rates of the three variational schemes with respect to the number of trigonometric terms used in the expansion $(M)$. Furthermore, it is of interest to investigate the convergence behaviors of the three variational principles under different compliance and density variations. Nemat-Nasser et al.\cite{nemat1975harmonic} in 1975 proved that the mixed quotient, in general, converges faster than the other quotients. Babuska and Osborn\cite{babuska1978numerical} presented in 1978 their theoretical analysis on the convergence rates of the three quotients, and related them to the function spaces of the density and compliance functions. They showed that when compliance, $D_{jkmn}$, is rough and density, $\rho$, is smooth, the stress Rayleigh quotient converges faster than the displacement Rayleigh quotient, however, when density is rough and compliance is smooth, the displacement Rayleigh quotient formulation results in faster convergence rates. In both situations the mixed quotient converges faster than the other two. In the following sections we explicitly calculate the convergence rate of the three quotients under different density and compliance variation.

\subsection{Convergence Rate}\label{amethod}

It can be shown that the error of any formulation considered here is bounded by the following inequality:
\begin{eqnarray}\label{errorboundM}
\vert \lambda_0 - \lambda_0^M \vert \leq CM^{-2\eta+\epsilon},
\end{eqnarray}
where $\lambda_0^M$ is the eigenvalue approximated by $M$ trigonometric terms. We omit the proof of the above inequality, (\ref{errorboundM}), and for details refer to Babuska and Osborn\cite{babuska1978numerical}. Since $\lambda=\omega^2$ and $f=w/2\pi$, frequency, $f$, should also be bounded by the above relation. We define the relative error as $\vert e\vert=\vert f_0 - f_0^M \vert/f_0$ and let $\xi=2\eta-\epsilon$. Taking natural $\log$ on (\ref{errorboundM}), we have
\begin{eqnarray}\label{ineqconvergentrate}
\log \vert e \vert \leq -\xi \log M +\log C,
\end{eqnarray}
where $C$ is the relative error when $M=1$ and $\xi$ is the relative convergence rate. By plotting $\log \vert e \vert$ as a function of $\log M$ we can extract the convergence rates of the different variational methods. 

\subsection{Special Cases: Either Density or Compliance is Constant}
According to Nemat-Nasser et al.\cite{nemat1975harmonic}, the mixed quotient (\ref{mixedvariational}) reduces to stress Rayleigh quotient (\ref{sigmaR}) under constant density and to the displacement Rayleigh quotient under constant compliance. We expect that the mix quotient converges as fast as the stress Rayleigh quotient in the former case and converges at the same rate as the displacement Rayleigh quotient in the latter case. The effect can be explained through the matrix form of the quotients. For instance, recasting (\ref{equationshomogeneousMatrix}) into the following form:
\begin{equation}\label{eigenvalueprobleminS}
\mathbf{H}^*\mathbf{\Omega}^{-1}\mathbf{H}\mathbf{S}=\lambda_{u\sigma}\mathbf{\Phi}\mathbf{S}
\end{equation}
Comparing to (\ref{eigenvalueMatrixstressR}), we need to show that $\mathbf{H}^*\mathbf{\Omega}^{-1}\mathbf{H}=-\bar{\boldsymbol{\rho}}$ when density is constant (since $\mathbf{\Phi}=\mathbf{D}$).
The matrix form of the mixed quotient formulation has the following coefficients:
\begin{eqnarray}\label{coefficients3dmixedQ}
\nonumber[\mathbf{H}]=i2\pi Q_k^{\theta\eta\xi}\int_\Omega \exp\left[i2\pi(Q_l^{\alpha\beta\gamma}-Q_l^{\theta\eta\xi})x_l\right]d\Omega\\
\nonumber[\mathbf{H}]^*=-i2\pi Q_n^{\alpha\beta\gamma}\int_\Omega \exp\left[i2\pi(Q_l^{\alpha\beta\gamma}-Q_l^{\theta\eta\xi})x_l\right]d\Omega\\
\nonumber[\mathbf{\Omega}]=\int_\Omega \rho(x_1,x_2,x_3)\exp\left[i2\pi(Q_l^{\alpha\beta\gamma}-Q_l^{\theta\eta\xi})x_l\right]d\Omega\\
\left[\mathbf{\Phi}\right]=\int_\Omega D_{jkmn}(x_1,x_2,x_3)\exp\left[i2\pi(Q_l^{\alpha\beta\gamma}-Q_l^{\theta\eta\xi})x_l\right]d\Omega, 
\end{eqnarray}
where $Q_l^{\alpha\beta\gamma}=T_{1l}(Q_1+\alpha)+T_{2l}(Q_2+\beta)+T_{3l}(Q_3+\gamma)$. 
The matrix form of the stress Rayleigh quotient formulation has the following coefficients:
\begin{eqnarray}\label{coefficients3dstressR}
\nonumber[\bar{\boldsymbol{\rho}}]=-4\pi^2 Q_n^{\alpha\beta\gamma}Q_k^{\theta\eta\xi}\int_\Omega \bar{\rho}(x_1,x_2,x_3)\exp\left[i2\pi(Q_l^{\alpha\beta\gamma}-Q_l^{\theta\eta\xi})x_l\right]d\Omega\\
\left[\mathbf{D}\right]=\int_\Omega D_{jkmn}(x_1,x_2,x_3)\exp\left[i2\pi(Q_l^{\alpha\beta\gamma}-Q_l^{\theta\eta\xi})x_l\right]d\Omega. 
\end{eqnarray}
When $\rho$ is constant, we have $\mathbf{H}\mathbf{\Omega}^{-1}\mathbf{H}^*=4\pi^2 Q_n^{\alpha\beta\gamma}Q_k^{\theta\eta\xi}\rho^{-1}\int_\Omega \exp\left[i2\pi(Q_l^{\alpha\beta\gamma}-Q_l^{\theta\eta\xi})x_l\right]d\Omega$, which is equal to (\ref{coefficients3dstressR})$^1$. Similarly, we can show that $\mathbf{H}\mathbf{\Phi}^{-1}\mathbf{H}^*=-\mathbf{C}$, when compliance is constant. In this case, the coefficients for displacement Rayleigh quotient formulation are:
\begin{eqnarray}\label{coefficients3ddispR}
\nonumber[\boldsymbol{\rho}]=\int_\Omega \rho(x_1,x_2,x_3)\exp\left[i2\pi(Q_l^{\alpha\beta\gamma}-Q_l^{\theta\eta\xi})x_l\right]d\Omega\\
\left[\mathbf{C}\right]=-4\pi^2 Q_n^{\alpha\beta\gamma}Q_k^{\theta\eta\xi}\int_\Omega C_{jkmn}(x_1,x_2,x_3)\exp\left[i2\pi(Q_l^{\alpha\beta\gamma}-Q_l^{\theta\eta\xi})x_l\right]d\Omega, 
\end{eqnarray}
where $C_{jkmn}$ is the material stiffness tensor satisfying $C_{jkmn}=D_{jkmn}^{-1}$. 

In summary, for the cases when either density or compliance is constant, the mixed quotient results in the same matrices as the stress Rayleigh quotient and the displacement Rayleigh quotient respectively. 

\section{Convergence rates for 1-D periodic composites}\label{1D}

\begin{figure}[htp]
\centering
\includegraphics[scale=.25]{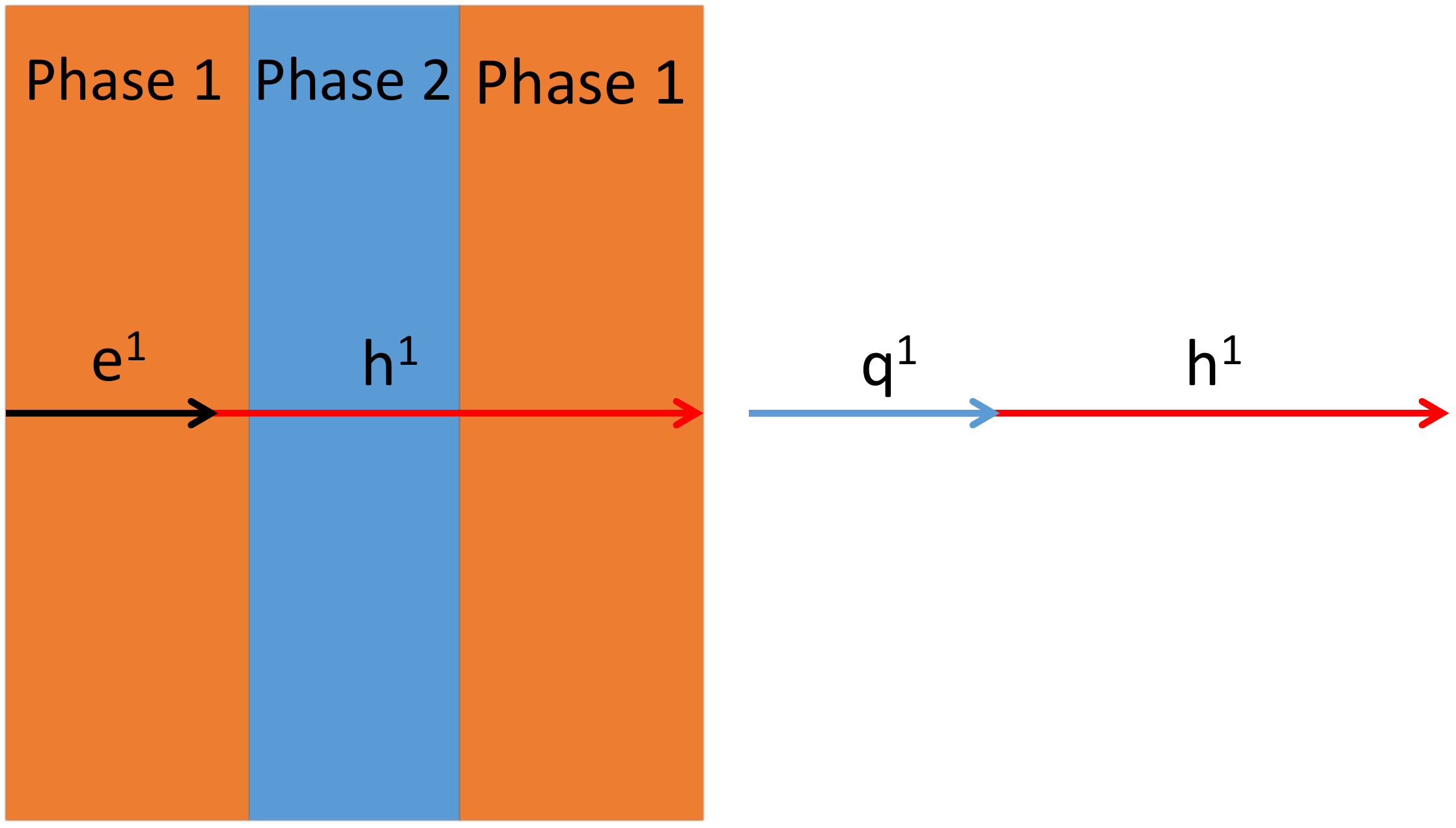}
\caption{Schematic of a 1-dimensional 2-phased periodic composite. The unit cell vector ($\mathbf{h}^1)$, reciprocal basis vector ($\mathbf{q}^1$), and the orthogonal vector ($\mathbf{e}^1$) are shown.}\label{1-DfVectors}
\end{figure}

There is only one possible Bravais lattice in 1-dimension with a unit cell vector whose length equals the length of the unit cell itself (Fig. \ref{1-DfVectors}). Without any loss of generality we take the direction of this vector to be the same as $\mathbf{e}^1$. If the length of the unit cell is $a$, then we have $\mathbf{h}^1=a\mathbf{e}^1$. The reciprocal vector is given by $\mathbf{q}^1=(2\pi/a)\mathbf{e}^1$. The wave-vector of a Bloch wave traveling in this composite is specified as $\mathbf{k}=Q_1\mathbf{q}^1$. To completely characterize the band-structure of the unit cell it is sufficient to evaluate the dispersion relation in the irreducible Brillouin zone ($-.5\leq Q_1\leq .5$). 

For plane longitudinal wave propagating in the $\mathbf{e}^1$ direction the only displacement component of interest is $u_1$ and the only relevant stress component is $\sigma_{11}$ (for plane shear waves traveling in $\mathbf{e}^1$ direction the quantities of interest are $u_2$ and $\sigma_{12}$). The equation of motion and the constitutive law are:
\begin{equation}\label{equationofmotion1D}
\sigma_{11,1}=-\lambda\rho(x_1) u_1; \quad \sigma_{11}=E(x_1)u_{1,1}
\end{equation}
where $E(x_1)$ is the spatially varying Young's modulus. The exact dispersion relation for 1-D longitudinal wave propagation in a periodic layered composite has been given by Rytov\cite{rytov1956acoustical},
\begin{eqnarray}\label{Rytov}
\nonumber \cos(\mathbf{k} a)=\cos(\omega h_1/c_1)\cos(\omega h_2/c_2)-\Gamma \sin(\omega h_1/c_1)\sin(\omega h_2/c_2),\\
\Gamma=(1+\kappa^2)/(2\kappa), \; \kappa=\rho_1c_1/(\rho_2c_2), 
\end{eqnarray}
where $h_i$ is the thickness, $\rho_i$ is the density, and $c_i$ is the longitudinal wave velocity of the $i$th layer $(i = 1,2)$ in a unit cell. We can solve for the corresponding wave number $\mathbf{k}$ by providing a range of frequency, $\omega$, using (\ref{Rytov}).

The displacement and stress fields are approximated by 1-D trigonometric functions:
\begin{equation}\label{approximation1d}
\bar{u}_1=\sum_{\alpha=-M}^MU^{\alpha}_1\exp\left[i(\mathbf{k}\cdot\mathbf{x}+2\pi\alpha H_1)\right],\quad \bar{\sigma}_{11}=\sum_{\alpha=-M}^MS^{\alpha}_{11}\exp\left[i(\mathbf{k}\cdot\mathbf{x}+2\pi\alpha H_1)\right]
\end{equation}
which can be further simplified to:
\begin{equation}\label{approximation1dS}
\bar{u}_1=\sum_{\alpha=-M}^MU^{\alpha}_1\exp\left[i2\pi(Q_1+\alpha)x_1/a\right],\quad \bar{\sigma}_{11}=\sum_{\alpha=-M}^MS^{\alpha}_{11}\exp\left[i2\pi(Q_1+\alpha)x_1/a\right],
\end{equation}
where $a$ is the periodicity length. 

\subsection{Details of the eigenvalue matrices}
For the displacement Rayleigh quotient the eigenvalue problem is:
\begin{eqnarray}\label{equationshomogeneous1DdispR}
\langle \left[E\bar{u}_{1,1}\right]_{,1}+\lambda_u\rho\bar{u}_1,f^\theta\rangle=0
\end{eqnarray}
where $-M\leq\alpha,\theta\leq M$. The above are transformed to the matrix form of (\ref{eigenvalueMatrixdispR}) with the following column vector:
\begin{eqnarray}\label{dispclmnvec}
\mathbf{U}=\{U^{-M}_1\;...\;U^{0}_1\;...\;U^{M}_1\}^T\nonumber\\
\end{eqnarray}
The associated coefficient matrices have the following elements:
\begin{eqnarray}\label{coefficients1ddispR}
\nonumber[\mathbf{C}]_{ij}=\frac{-4\pi^2(Q_1+i-M-1)(Q_1+j-M-1)}{a^2}\int_0^aE(x_1)\exp\left[i2\pi(i-j)x_1/a\right]dx_1\\
\nonumber[\boldsymbol{\rho}]_{ij}=\int_0^a\rho(x_1)\exp\left[i2\pi(i-j)x_1/a\right]dx_1\\
i,j=1,2,...(2M+1)
\end{eqnarray}

For the stress Rayleigh quotient the eigenvalue problem is:
\begin{eqnarray}\label{equationshomogeneous1DstressR}
\langle\left[\bar{\rho}\bar{\sigma}_{11,1}\right]_{,1}+\lambda_\sigma D\bar{\sigma}_{11},f^{\theta}\rangle=0
\end{eqnarray}
where $-M\leq\alpha,\theta\leq M$ and $D=1/E$. The above are transformed to the matrix form of (\ref{eigenvalueMatrixstressR}) with the following column vector:
\begin{eqnarray}\label{stressclmnvec}
\mathbf{S}=\{S^{-M}_{11}\;...\;S^{0}_{11}\;...\;S^{M}_{11}\}^T\nonumber\\
\end{eqnarray}
The associated coefficient matrices have the following elements:
\begin{eqnarray}\label{coefficients1dstressR}
\nonumber[\mathbf{D}]_{ij}=\int_0^aD(x_1)\exp\left[i2\pi(i-j)x_1/a\right]dx_1\\
\nonumber[\boldsymbol{\bar{\rho}}]_{ij}=\frac{-4\pi^2(Q_1+i-M-1)(Q_1+j-M-1)}{a^2}\int_0^a\bar{\rho}(x_1)\exp\left[i2\pi(i-j)x_1/a\right]dx_1\\
i,j=1,2,...(2M+1). 
\end{eqnarray}

For the mixed-quotient the eigenvalue problem is:
\begin{eqnarray}\label{equationshomogeneous1DmixedQ}
\langle\bar{\sigma}_{11,1}+\lambda_{u\sigma}\rho\bar{u_1},f^{\theta}\rangle=0\nonumber\\
\langle D\bar{\sigma}_{11}-\bar{u}_{1,1},f^{\theta}\rangle=0
\end{eqnarray}
where $-M\leq\alpha,\theta\leq M$. The above are transformed to the matrix form of (\ref{equationshomogeneousMatrix}) with column vectors (\ref{dispclmnvec}) and (\ref{stressclmnvec}).
The associated coefficient matrices have the following elements:
\begin{eqnarray}\label{coefficients1dmixedQ}
\nonumber[\mathbf{H}]_{ij}=\frac{i2\pi(Q_1+i-M-1)}{a}\int_0^a\exp\left[i2\pi(i-j)x_1/a\right]dx_1\\
\nonumber[\mathbf{\Omega}]_{ij}=\int_0^a\rho(x_1)\exp\left[i2\pi(i-j)x_1/a\right]dx_1\\
\nonumber[\mathbf{\Phi}]_{ij}=\int_0^aD(x_1)\exp\left[i2\pi(i-j)x_1/a\right]dx_1\\
\nonumber\left[\mathbf{H}\right]^*_{ij}=\frac{-i2\pi(Q_1+j-M-1)}{a}\int_0^a\exp\left[i2\pi(i-j)x_1/a \right]dx_1\\
i,j=1,2,...(2M+1). 
\end{eqnarray}

Now the eigenvalue problems can be solved for the frequencies $\omega$ which correspond to an assumed value of $Q_1$.

\subsection{1-D 2-phase Layered Composite: Comparison with exact Solution }\label{2layercomp}

The exact solution for wave propagation in 1-D for a 2-phase periodic composite was first given in (\ref{Rytov}). Here we present a comparison of the results from the displacement Rayleigh quotient, stress Rayleigh quotient, and mixed quotient formulations with the exact Rytov solution. We have not employed numerical integration in the following results because for the 1-D case the integrals in (\ref{coefficients1ddispR}), (\ref{coefficients1dstressR}) and (\ref{coefficients1dmixedQ}) can be calculated exactly.

\begin{figure}[htp]
\centering
\includegraphics[scale=.5]{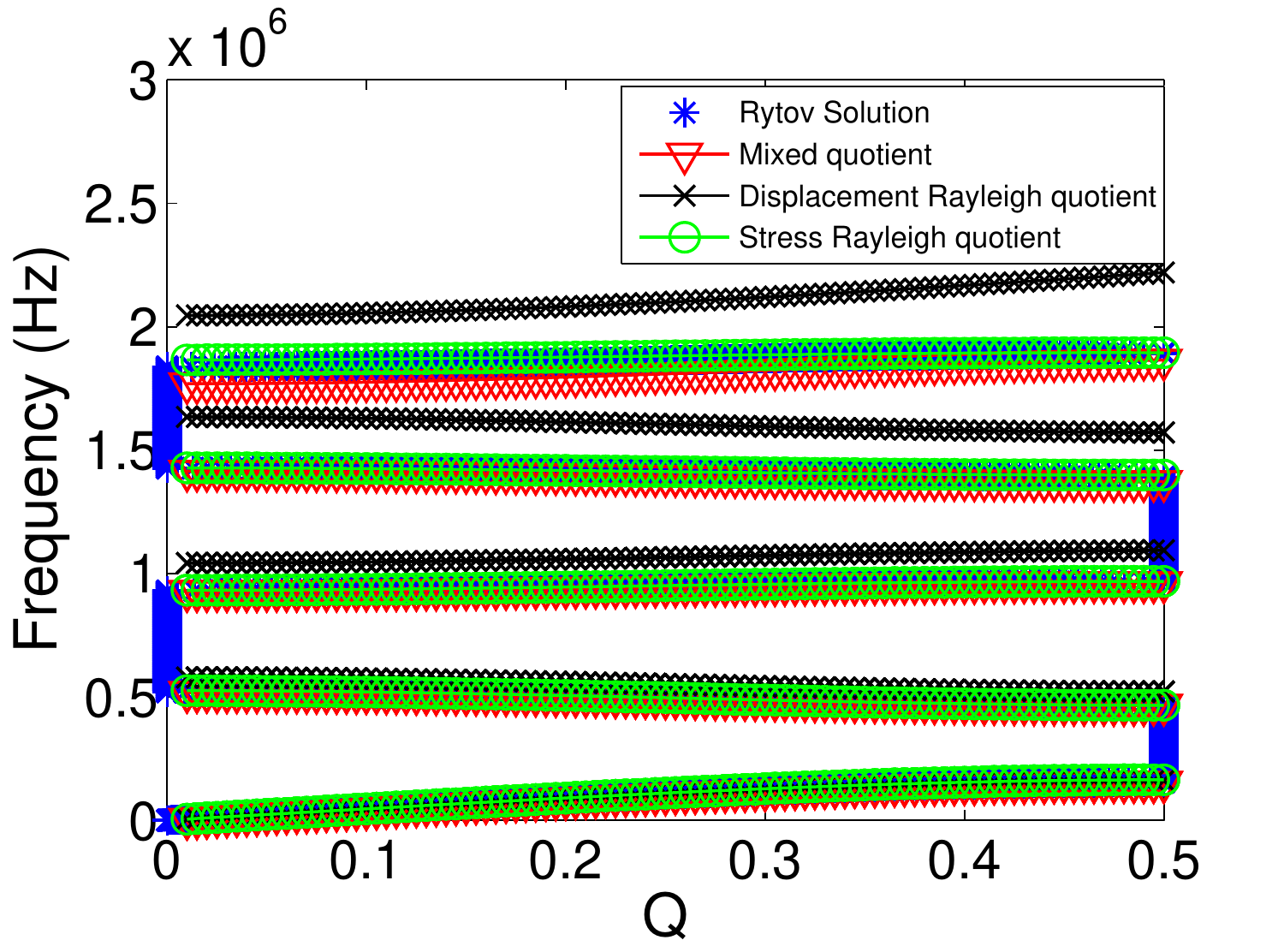}
\caption{Comparison of the displacement Rayleigh quotient, stress Rayleigh quotient and mixed quotient formulation results with exact solution. The first 5 branches are shown for $M=4$.}\label{comparewithexact}
\end{figure}

The composite under consideration is a 2-phase layered composite consisting of the following 2 phases:
\begin{enumerate}
\item Phase 1: $E_1=8GPa,\; \rho_1=1000kg/m^3,\; thickness=0.003m$

\item Phase 2: $E_2=300GPa,\; \rho_2=8000kg/m^3,\; thickness=0.0013m$ 
\end{enumerate}
Fig. (\ref{comparewithexact}) shows the comparison of the results from the displacement Rayleigh quotient, stress Rayleigh quotient and mixed quotient formulations with the Rytov solution for the first five branches. The results are calculated for $0\leq Q_1 \leq 0.5$ at a total of 99 points. The variational results are shown for $M=4$ for each variational scheme. 
It is clear from Fig. (\ref{comparewithexact}) that the mixed quotient and the stress Rayleigh quotient capture the first 3 branches very well. In general, more trigonometric expansion terms are required for higher accuracy in higher branches and the accuracy of all three quotients worsens as we consider higher branches. For the 4th and 5th branches, the stress Rayleigh quotient gives better approximation than the mixed quotient, however it shows slower convergence (Table \ref{errorCexact}). 
\begin{table}[htp]
\caption{The average convergence rates, $\xi$, and the initial errors, $C$, approximated using the results from $M=2$ to $15$ for the first five branches.}\label{errorCexact}
\centering
\begin{tabular}{lccccc}
\hline $\xi$ & branch 1& branch 2& branch 3& branch 4& branch 5\\
\hline Mixed quotient & 0.8173 & 2.2927 & 2.8271 & 3.1052 &	3.4739\\
Stress Rayleigh quotient & 1.0067 &	0.9495 & 1.0209 & 1.6483 & 1.7887 \\
Displacement Rayleigh quotient & 0.9840 & 0.9687 & 1.2320 & 1.3545 & 1.4206 \\
\hline $C$ & branch 1& branch 2& branch 3& branch 4& branch 5\\
\hline Mixed quotient & 1.87E-4 & 1.44E-2 & 0.1356 & 0.7785 & 3.0386 \\
Stress Rayleigh quotient & 0.2547 & 2.46E-2 & 8.61E-3 & 3.74E-2 & 0.1636 \\ 
Displacement Rayleigh quotient & 0.2203 & 0.4370 & 	0.8348 & 1.1308 & 1.2510\\
\hline
\end{tabular}
\end{table}

\subsection{Convergence under Different Density and Compliance Distributions}
Convergence of the variational schemes depends upon the roughness of the density and compliance distributions over the unit cell. Studying the convergence behavior of the three methods under arbitrary variations of compliance and density is complicated by the absence of exact solutions to such problems. We tackle this problem by noting that the result of the mixed quotient formulation using sufficient number of trigonometric terms can serve as a good enough approximation to the exact solution, especially for lower branches. For 1-D calculations, we use the mixed variation result at $M=50$ to serve as the reference solution, for which the relative error on the first branch, comparing to Rytov's solution, is under $3.5\times10^{-3}\%$. The convergence rates are calculated using the results from $M=2$ to $15$.
\begin{figure}[htp]
\centering
\includegraphics[scale=.5]{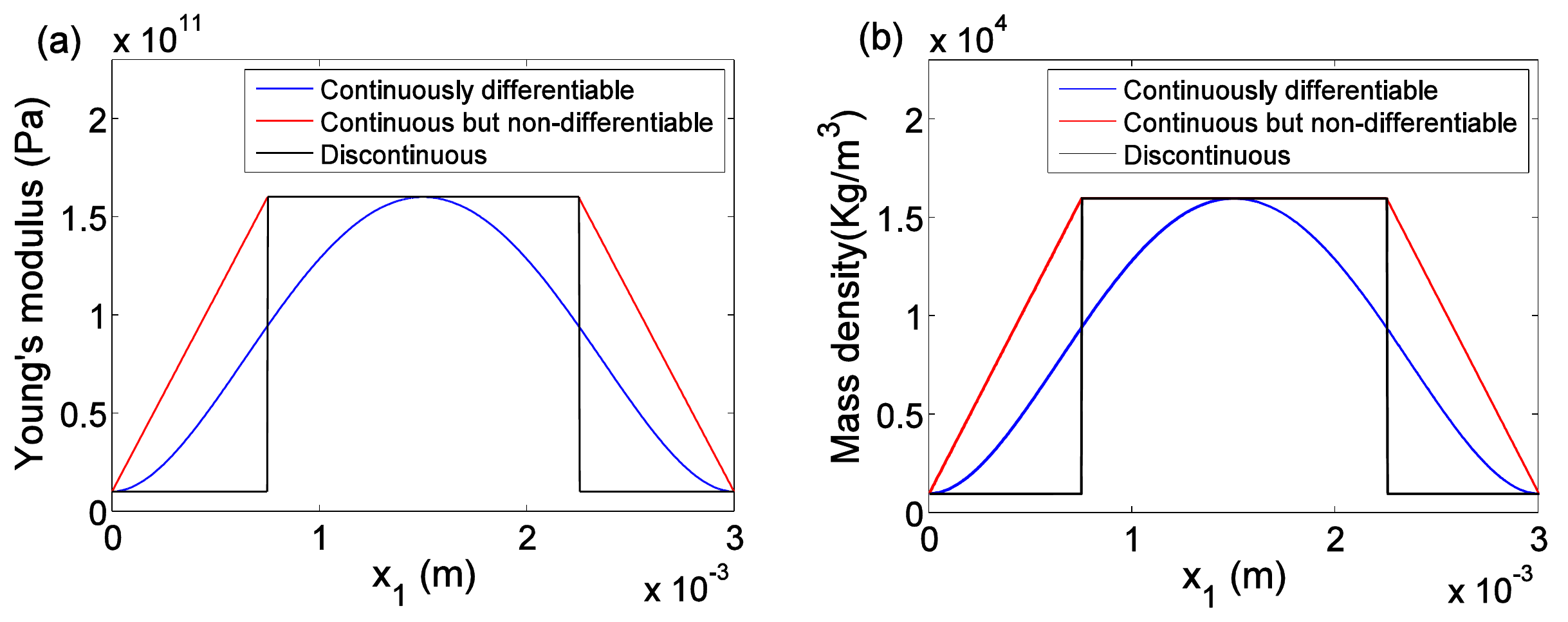}

\caption{(a) For the cases where density $\rho$ is constant, we apply different Young's modulus distributions in an unit cell, where $E_{max}/E_{min}=16$. (b) For the cases where compliance $D$ is constant, we apply different density distributions in a unit cell, where $\rho_{max}/\rho_{min}=16$.}\label{materialdistr}
\end{figure}

We calculate cases where either density is constant ($\rho=1000kg/m^3$), or Young's modulus is constant ($E=10Gpa$). The corresponding Young's modulus and density distributions are given in Fig. (\ref{materialdistr}). As shown in Fig. (\ref{materialdistr}), for the continuously differentiable cases, the Young's modulus and density distribution functions are of 4th order and their derivatives at the boundaries of the unit cell are zero. As compliance $D=1/E$, $D$ is also continuously differentiable given that $dD/dx=d/dx(1/E)=-(1/E^2)dE/dx$. For the continuous but non-differentiable cases, the material properties are changing linearly in one phase and are constant in the central phase. For the discontinuous cases, the material properties have large contrasts at the interfaces.  

\subsubsection{Convergence rates under different compliance distributions}

\begin{figure}[htp]
\centering
\includegraphics[scale=.5]{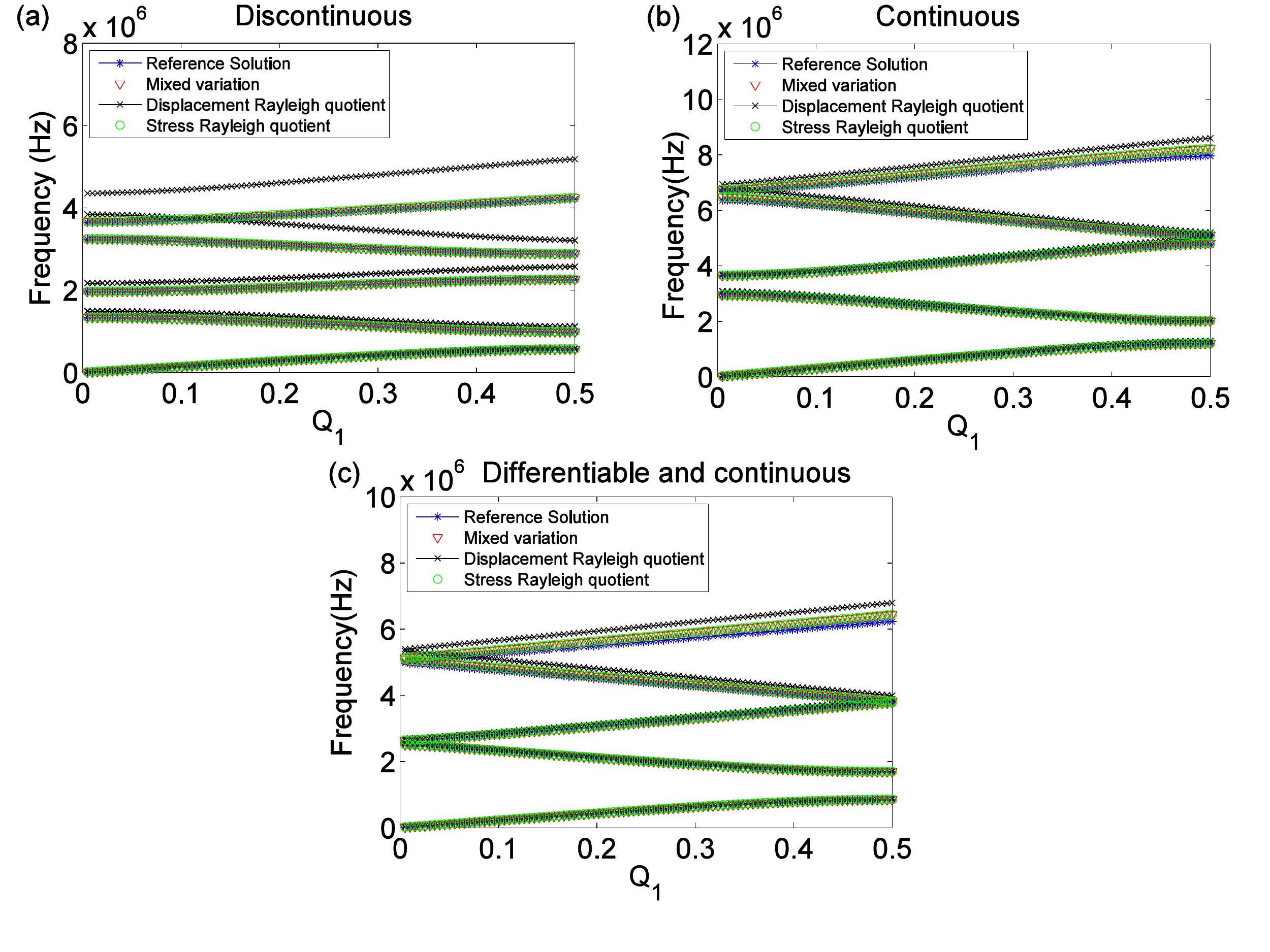}

\caption{Bandstructure of 1-D periodic constant density composite with (a)discontinuous , (b)continuous but non-differentiable and (c)continuously differentiable compliance distributions.}\label{constantrho}
\end{figure}

\begin{figure}[htp]
\centering
\includegraphics[scale=.5]{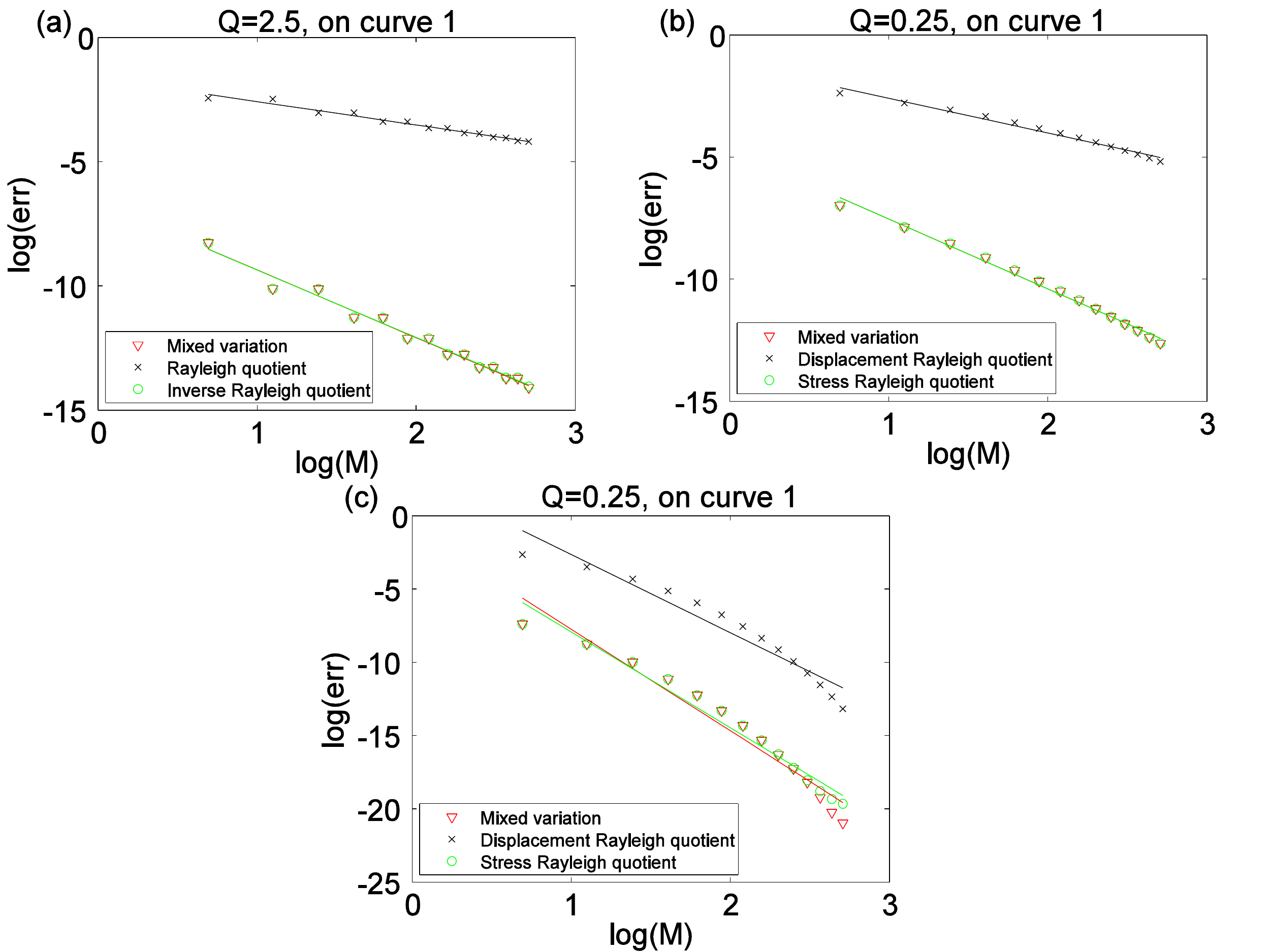}

\caption{The log(err)-log(M) plot for (a) Discontinuous, (b) Continuous but non-differentiable,  (c) Continuously differentiable compliance variations (constant density).}\label{constantrholog}
\end{figure}

Fig. (\ref{constantrho}) shows the bandstructure calculations for the different compliance distributions discussed in Fig. (\ref{materialdistr}a). It can be seen that the solutions for different quotients overlap on one another for the lower branches. As frequency increases, noticeable  errors emerge but the curves of the mixed quotient overlap the curves of the stress Rayleigh quotient.
Fig. (\ref{constantrholog}) shows the $\log(err)-\log(M)$ plot for the three cases at $Q=0.25$, on the first branch. The slope of the fitting lines in each case is the average relative convergence rate. It can be seen that the results of the mixed quotient overlap with the stress Rayleigh quotient suggesting that they have the same convergence rate and that the slope of displacement Rayleigh quotient fitting line is less steep than the others showing that it converges slower. 

Table (\ref{averageconvrconstrho}) gives the average convergence rates for the three variational formulations calculated over different branches for all three cases. In this constant density case, $\xi_{M}\approx\xi_{I}$, thus verifying that the mixed quotient converges as fast as the stress Rayleigh quotient in this situation. Furthermore, it is clear that the convergence rates of the three quotients for the continuously differentiable case are larger than the other two cases considered. In general, the convergence rates are seen to become larger as frequency increases for the discontinuous case, however, this is not generally true for the other two cases. The relative errors of the mixed quotient and the displacement Rayleigh quotient $\vert e \vert _M/\vert e \vert_R$ for the three cases considered here are proportional to $M^{-1.84}$, $M^{-1.32}$, $M^{-1.52}$ respectively.
\begin{table}[htp]
\caption{Convergence rates, $\xi$, on five dispersion curves measured at $Q=0.25$ for discontinuous, continuous but non-differentiable and continuously differentiable compliance variations (density constant). The subscripts $M$, $R$, $I$ refer the mixed quotient, the displacement Rayleigh quotient and stress Rayleigh quotient respectively. }\label{averageconvrconstrho}
\centering
\begin{tabular}{llllllllll}
\hline $Q=0.25$& \multicolumn{3}{l}{Discontinuous} & \multicolumn{3}{l}{Continuous,non-differentiable}& \multicolumn{3}{l}{Continuously differentiable}\\
\cline{2-4}  \cline{5-7}  \cline{8-10}
  & $\xi_{M}$ & $\xi_{R}$ & $\xi_{I}$ & $\xi_{M}$ & $\xi_{R}$ & $\xi_{I}$ & $\xi_{M}$ & $\xi_{R}$ & $\xi_{I}$\\
\hline
Curve 1 & 2.7284 & 0.9404 & 2.7079 & 2.8571 & 1.4214 & 2.8569 & 6.9244 & 5.3303 & 6.5389\\
Curve 2 & 2.7893 & 1.1991 & 2.7873 & 3.0222 & 1.6177 & 3.0223 & 6.8564 & 5.2507 & 6.9125\\
Curve 3 & 3.1230 & 1.2897 & 3.1225 & 3.0154 & 1.7418 & 3.0154 & 6.5615 & 5.0656 & 6.5583\\
Curve 4 & 3.7763 & 1.6516 & 3.7761 & 3.2786 & 1.9673 & 3.2786 & 6.3230 & 4.8121 & 6.3230\\
Curve 5 & 4.0255 & 2.1819 & 4.0254 & 3.1347 & 1.9668 & 3.1347 & 5.9079 & 4.5305 & 5.9080\\
\hline
\end{tabular}
\end{table}
 
\subsubsection{Convergence rates under different density distributions}

We now consider the cases of constant compliance and variable density as described in Fig. (\ref{materialdistr}b). Table (\ref{averageconvrconstE}) gives the average convergence rates over the first five branches ($Q=0.25$) for the three variational methods when compliance is constant. In this constant compliance case, $\xi_{M}\approx\xi_{R}$, thus verifying that the mixed quotient converges as fast as the displacement Rayleigh quotient in this situation. Furthermore, it is clear that the convergence rates of the three quotients for the continuously differentiable case are larger than the other two cases considered. In general, the convergence rates are seem to become larger as frequency increases. The relative errors of the mixed quotient and the stress Rayleigh quotient $\vert e\vert_M/\vert e\vert_I$ for the three cases considered here are proportional to $M^{-1.88}$, $M^{-1.95}$, $M^{-2.35}$ respectively.
\begin{table}[htp]
\caption{Convergence rates, $\xi$, on five dispersion curves measured at $Q=0.25$ for discontinuous, continuous but non-differentiable and continuously differentiable density variations (compliance constant). The subscripts $M$, $R$, $I$ refer the mixed quotient, the displacement Rayleigh quotient and stress Rayleigh quotient respectively. }\label{averageconvrconstE}
\centering
\begin{tabular}{llllllllll}
\hline $Q=0.25$ &\multicolumn{3}{l}{Discontinuous} & \multicolumn{3}{l}{Continuous,non-differentiable}& \multicolumn{3}{l}{Continuously differentiable}\\
\cline{2-4}  \cline{5-7}  \cline{8-10}
& $\xi_{M}$ & $\xi_{R}$ & $\xi_{I}$ & $\xi_{M}$ & $\xi_{R}$ & $\xi_{I}$ & $\xi_{M}$ & $\xi_{R}$ & $\xi_{I}$\\
\hline
Curve 1 & 2.7805& 2.7805& 0.9388 & 4.2087&	4.2088&	2.4163 & 7.1648& 7.1872& 6.6472\\
Curve 2 & 2.7863& 2.7863& 1.1273 & 4.9232& 4.9232& 2.9175 & 9.8039& 9.7735& 7.0920\\
Curve 3 & 3.0680& 3.0680& 1.1583& 4.7126& 4.7126& 2.8078&10.819& 10.801& 8.0093\\
Curve 4 & 3.7986& 3.7986& 1.7437& 5.8581& 5.8581& 3.7621& 13.213& 13.208& 10.399\\
Curve 5 & 4.0133&	4.0133&	2.0724& 5.6898&	5.6898&	3.7535&	13.930& 13.926& 11.012\\
\hline
\end{tabular}

\end{table}

\section{2-D periodic composites}\label{2D}
There are five possible Bravais lattices in 2 dimensions. However, they can be specified using two unit cell vectors $(\mathbf{h}^1,\mathbf{h}^2)$. The reciprocal vectors are $\mathbf{q}^1$, $\mathbf{q}^2$. The wave-vector of a Bloch-wave traveling in this composite is specified as $\mathbf{k}=Q_1\mathbf{q}^1+Q_2\mathbf{q}^2$. To characterize the band-structure of the unit cell we evaluate the dispersion relation along the boundaries of the irreducible Brillouin zone $(0\leq Q_1\leq .5, \; Q_2=0; \; Q_1=.5, \; 0\leq Q_2\leq .5; \; 0\leq Q_1\leq .5, Q_2=Q_1)$. In traditional notation these boundaries are specified as $\Gamma-X-M-\Gamma$.

For the purpose of demonstration and comparison we consider the case of plane-strain state in the composite. The relevant stress components for the plane-strain case are $\sigma_{11}$, $\sigma_{22}$, $\sigma_{12}$ and the relevant displacement components are $u_1$, $u_2$. The equations of motion and the constitutive relation are,
\begin{eqnarray}\label{equationofmotion2D}
\nonumber \sigma_{jk,k}=-\lambda \rho(\mathbf{x})u_j,\; D_{jkmn}(\mathbf{x})\sigma_{mn}=u_{j,k}\\
j,k,m,n=1,2,
\end{eqnarray}
where $\mathbf{D}$ is the compliance tensor. For an isotropic material in plane strain, $\mathbf{D}$ is given by,
\begin{eqnarray}\label{compliancetensor}
\nonumber D_{jkmn}=\dfrac{1}{2\mu}\left[\dfrac{1}{2}\left(\delta_{jm}\delta_{kn}+\delta_{jn}\delta_{km}\right)-\dfrac{\lambda}{2\left(\mu+\lambda\right)}\delta_{jk}\delta_{mn}\right]\\
j,k,m,n=1,2,
\end{eqnarray}
where $\lambda$, $\mu$ are the Lam{\'e} constants. The stresses and displacements are approximated by the following 2-D periodic functions:
\begin{equation}\label{approximation2D}
\bar{u}_j=\sum_{\alpha,\beta=-M}^{M}U^{\alpha\beta}_j \exp\left[i2\pi Q^{\alpha\beta}_l x_l\right],\quad \bar{\sigma}_{jk}=\sum_{\alpha,\beta=-M}^{M} S^{\alpha\beta}_{jk}\exp\left[i2\pi Q^{\alpha\beta}_l x_l\right],
\end{equation}
where
\begin{eqnarray}
\nonumber Q^{\alpha\beta}_1=T_{11}(Q_1+\alpha)+T_{21}(Q_2+\beta)\\
Q^{\alpha\beta}_2=T_{12}(Q_1+\alpha)+T_{22}(Q_2+\beta),
\end{eqnarray}
and the square matrix $[\mathbf{T}]$ is the inverse of the matrix $[\mathbf{A}]$ with components $[\mathbf{A}]_{jk}=\mathbf{h}^j\cdot\mathbf{e}^k$. We omit here the details of the formulation of the matrices for solving the eigenvalue problem of the 2-D composites. (See Srivastava and Nemat-Nasser\cite{srivastava2014mixed} for details.)
To calculate integrals, such as in (\ref{coefficients3dmixedQ}, \ref{coefficients3dstressR}, \ref{coefficients3ddispR}), we have employed numerical integrals over automatically generated subdomains of $\Omega$. A total 2402 triangular subdomains of an unit cell are generated by the freely available FE software GMSH\cite{geuzaine2009gmsh}. This process is executed using GPU computation and the eigenvalue problem is finally calculated in Python environment. 

\begin{figure}[htp]
\centering
\includegraphics[scale=.35]{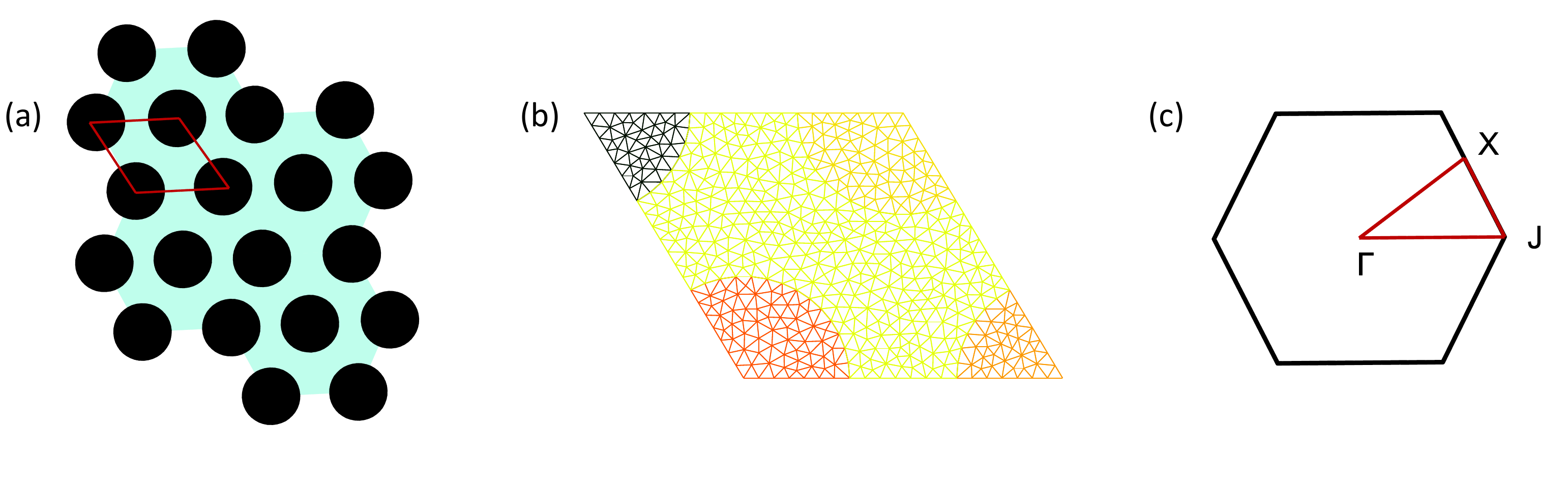}

\caption{(a) Schematic of the 2-D periodic composite made from steel cylinders distributed in hexagonal packing in epoxy matrix; (b) discretization of the unit cell; (c) irreducible Brillouin zone in the reciprocal lattice.}\label{2dschematic}
\end{figure}
The example unit cell considered in the following calculation is a 2-phase hexagonal unit cell (Fig. (\ref{2dschematic})). The diameter of the inclusion is $4mm$ and the lattice constant is $6.023mm$ leading to a filling ratio of $40\%$. The material properties we applied are $E_{steel}=207.1475GPa$, $\nu_{steel}=0.2786$, $\rho_{steel}=7780kg/m^3$, $E_{epoxy}=4.0785GPa$, $\nu_{epoxy}=0.3779$, $\rho_{epoxy}=1142kg/m^3$. 
\begin{figure}[htp]
\centering
\includegraphics[scale=.3]{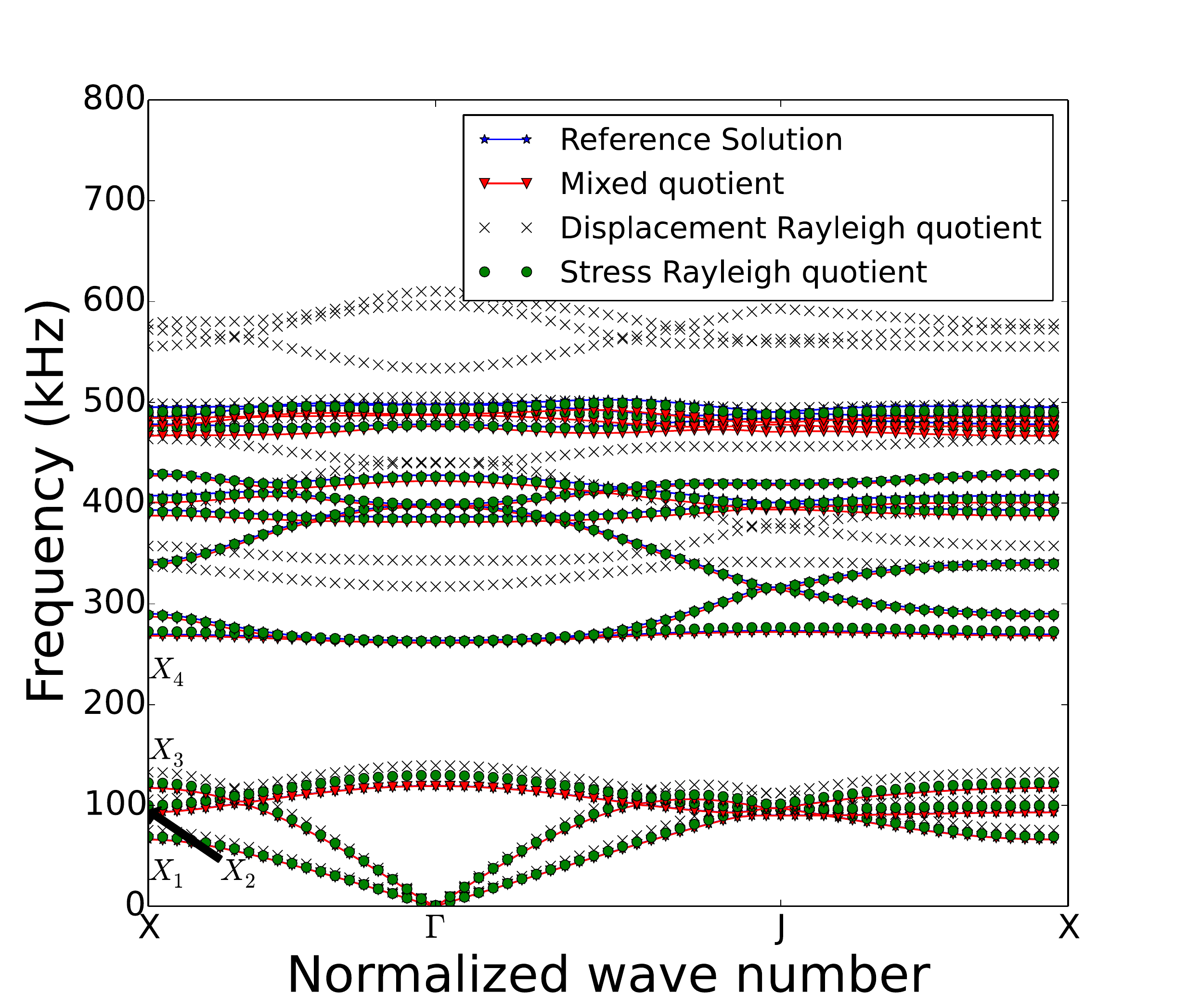}
\caption{Bandstructure of the three variational principles using 162 terms ($M=4$). }\label{compare2D}
\end{figure}

The mixed quotient results using a total of 3362 trigonometric terms ($M=20$) serve as the reference solution. There are two complete bandgaps between $119-263KHz$, $429-475KHz$ and they are in very good agreement with the PWE calculations provided by Vasseur et al.\cite{vasseur2001experimental}. For the bandstructure results shown in Fig. (\ref{compare2D}). The IBZ is discretized at 64 points and a total of 162 terms in the Fourier expansion ($M=4$) are used. It is clear that the mixed quotient and stress Rayleigh quotient match well with the reference solution while the displacement Rayleigh quotient shows considerable errors as frequency increases. Due to the low density contrast and high compliance contrast, the stress Rayleigh quotient performs much better than the displacement Rayleigh quotient.   
\begin{figure}[htp]
\centering
\includegraphics[scale=.3]{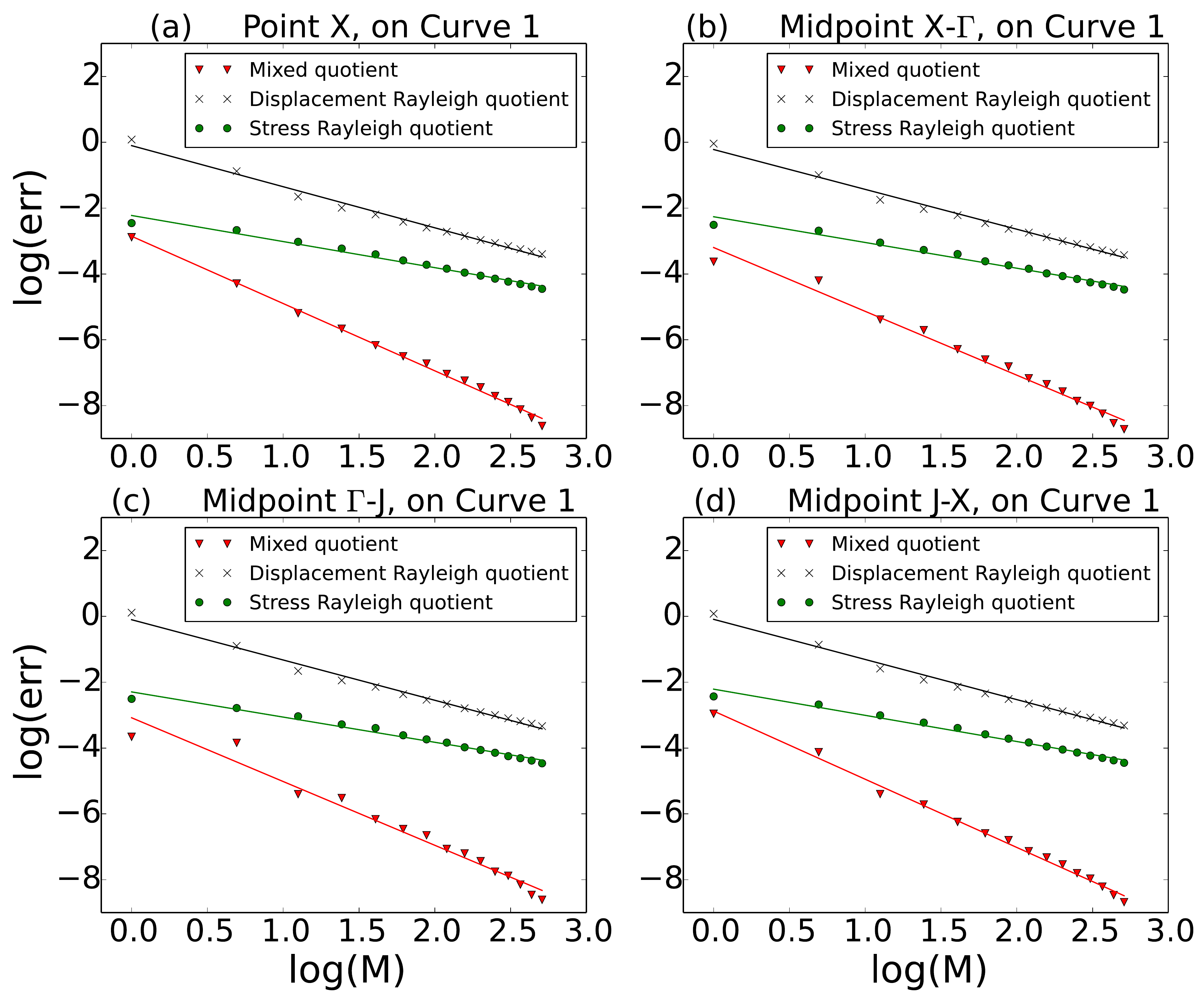}

\caption{$\log(err)-\log(M)$ plot for the wavevector (a) on point $X$, (b) on the midpoint between $X-\Gamma$, (c) on the midpoint between $\Gamma-J$, (d) on the midpoint between $J-X$.}\label{2Dlog}
\end{figure}
We are interested in the convergence rates of the three methods applying to the 2-D composites bandstructure calculation. The convergence rates are calculated for the high symmetry point $X$ and the midpoints between $X-\Gamma$, $\Gamma-J$ and $J-X$. Fig. (\ref{2Dlog}) shows the $\log(err)-\log(M)$ plot of the first branch for $M=1$ to $15$. The slope of the mixed quotient fitting line is steeper than the other two methods suggesting that it displays higher convergence. The stress Rayleigh quotient has smaller slope than the displacement Rayleigh quotient, signifying slower convergence. However, the stress Rayleigh quotient shows a smaller initial error than the displacement Rayleigh quotient. The convergence rates on the first 8 curves are given in Table (\ref{averageconvr2D}). It can be seen that the convergence rates in the 2-D case generally have the same characteristics as 1-D. The mixed quotient generally converges faster than the other two methods. In some cases (5th and 7th curves) the stress Rayleigh quotient is seen to converge slightly faster than the mixed quotient. For lower curves, all three methods converge to good results by using a small number of expansion terms but only the mixed quotient maintains a high convergence rate as shown in Fig. (\ref{2Dlog}), suggesting that the mixed quotient is more useful when computing high precision results.

\begin{table}[htp]
\caption{This table includes the values of average convergence rates, $\xi$, from curve 1 to curve 8 measured at $X$ and the midpoints of $X-\Gamma$, $\Gamma-J$, $J-X$. The subscription $M$  means mixed quotient, $R$ stands for displacement Rayleigh quotient and stress Rayleigh quotient is denoted by $I$. }\label{averageconvr2D}
\centering
\begin{tabular}{lccc|lccc}
\hline Curve 1  & $\xi_{M}$ & $\xi_{R}$ & $\xi_{I}$ & 	Curve 2  & $\xi_{M}$ & $\xi_{R}$ & $\xi_{I}$\\
\hline $X$& 2.0434& 1.2477& 0.7921& $X$& 2.1401& 1.2537& 0.9254\\
$X-\Gamma$&	1.9384&	1.2103&	0.7818& $X-\Gamma$&	1.8682&	1.1770&	0.8552\\
$\Gamma-J$&	1.9363&	1.2232&	0.7649& $\Gamma-J$&	1.7762&	1.1246&	0.8642\\
$J-X$& 2.0715& 1.2191& 0.7950& $J-X$& 2.1718& 1.2332& 0.9219\\ 
\hline Curve 3 & $\xi_{M}$ & $\xi_{R}$ & $\xi_{I}$ & Curve 4 & $\xi_{M}$ & $\xi_{R}$ & $\xi_{I}$\\
\hline$X$& 2.2317& 1.1659& 0.8810& $X$& 2.6812& 1.1478& 0.6027\\
$X-\Gamma$&	2.3124&	1.2572&	0.9244& $X-\Gamma$&	2.4236&	1.1812&	0.8059\\
$\Gamma-J$&	2.2346&	1.2660&	0.9154& $\Gamma-J$&	2.4417&	1.1836&	0.8048\\
$J-X$& 2.1950& 1.1920& 0.8627& $J-X$&	2.6764&	1.1336&	0.0113\\
\hline Curve 5  & $\xi_{M}$ & $\xi_{R}$ & $\xi_{I}$ & 	Curve 6  & $\xi_{M}$ & $\xi_{R}$ & $\xi_{I}$\\
\hline $X$& 2.2626& 1.2946& 2.0239& $X$& 2.5317& 1.3958& 2.2040\\
$X-\Gamma$&	2.2276&	1.2492&	2.3409& $X-\Gamma$&	2.7772&	1.3047&	2.1939\\
$\Gamma-J$&	2.2504&	1.2557&	2.3309& $\Gamma-J$&	2.7225&	1.3128&	2.2926\\
$J-X$& 2.3007& 1.3154& 2.5746& $J-X$& 2.4972& 1.4046& 2.9608\\
\hline Curve 7 & $\xi_{M}$ & $\xi_{R}$ & $\xi_{I}$ & Curve 8 & $\xi_{M}$ & $\xi_{R}$ & $\xi_{I}$\\
\hline $X$& 2.6452& 1.0470& 2.6516& $X$& 2.4576& 1.2031& 2.2978\\
$X-\Gamma$&	2.4886&	0.9285&	2.9138& $X-\Gamma$&	2.9559&	1.4556&	1.1620\\
$\Gamma-J$&	2.5141&	0.9347&	2.9198& $\Gamma-J$&	2.9970&	1.4597&	0.7055\\
$J-X$& 2.6273& 1.0146& 2.7722& $J-X$&	2.4676&	1.1923&	2.3877\\
\hline
\end{tabular}
\end{table}
\section{Comparison with FEM}
The variation formulations of this elastodynamic problem remain the same regardless of test functions which expand the displacement and stress fields. Thus, the minimization problems mentioned in the previous section can be solved by using natural basis functions. Displacement Rayleigh quotient method in natural basis is equivalent to the formulation for displacement based FEM (see (\ref{equationshomogeneousDispR})). It is, therefore, interesting to compare the convergence of the mixed quotient method with FEM using the same basis function. In this case, we choose $f^{\alpha\beta\gamma}$ to be second order Lagrange basis functions, let $U^{\alpha\beta\gamma}_j=\bar{U}^{\alpha\beta\gamma}_j\exp[i\mathbf{k\cdot x}]$, $S^{\alpha\beta\gamma}_{jk}=\bar{S}^{\alpha\beta\gamma}_{jk}\exp[i\mathbf{k\cdot x}]$ and force the reduced displacement, $\bar{U}$, and stress, $\bar{S}$, fields to satisfy the periodic boundary conditions. 

Since the mixed quotient (\ref{mixedvariational}) remains the same for Lagrange basis functions, it is easy to derive the matrix form from (\ref{equationshomogeneous}) and obtain
\begin{eqnarray}\label{LagrangeVariationMatrix}
\nonumber \mathbf{E\bar{S}}-\lambda\mathbf{\Omega \bar{U}}=0\\
\mathbf{\Phi \bar{S}}-\mathbf{F\bar{U}}=0.
\end{eqnarray}
Notice the exponential terms are canceled in the system of equations, therefore, Bloch boundary conditions are reduced to periodic boundary condition. $\mathbf{E}$ and $\mathbf{F}$ come from the derivatives of displacement and stress. They can be expressed as 
\begin{eqnarray}
\nonumber \mathbf{E}=\mathbf{A}^*-\mathbf{B} \\
\nonumber\mathbf{F}=\dfrac{1}{2}(\mathbf{A}+\mathbf{A}^*+\mathbf{B}+\mathbf{B}^*)\\
\nonumber\mathbf{A}=\int_\Omega f^{\alpha\beta\gamma}_{,j}f^{\theta\eta\xi}d\Omega \\
\mathbf{B}=\int_\Omega ik_jf^{\alpha\beta\gamma}f^{\theta\eta\xi}d\Omega,
\end{eqnarray}
where $f_{,j}^{\alpha\beta\gamma}$ is the $j$th derivative of Lagrange basis function $f^{\alpha\beta\gamma}$, $k_j$ is the $j$th component of wave vector $\mathbf{k}$. The integrals above are calculated using Gaussian quadrature rule. The general eigenvalue form of mixed variation using Lagrange elements can be written as 
\begin{equation}
\mathbf{E \Phi^{-1} F\bar{U}}=\lambda \mathbf{\Omega \bar{U}}
\end{equation}

In order to implement FEM, the system of equations are usually written in terms of displacement field, such as (\ref{equationofmotionDisp}). By taking the inner product of the chosen test function  and (\ref{equationofmotionDisp}) over the the whole solution domain, we obtain the following general eigenvalue problem:
\begin{equation}\label{eigenFEM}
\left(\mathbf{K(k)}+\lambda\mathbf{M}\right)\mathbf{\bar{U}}=0, 
\end{equation}
where $\mathbf{K(k)}$ is the global stiffness matrix associated with wave vector $\mathbf{k}$, and $\mathbf{M}$ is the global mass matrix. The 1-D 2-phase composite material properties for which the following comparison results are presented are given in Section \ref{2layercomp}.

The 1-D domain is decomposed into two subdomains based on the material properties and discretized into $N$ elements in total, which results in matrices size of $(2N+1)\times(2N+1)$ for both methods. The convergence rates are calculated in terms of number of elements, because the increasing discretization leads to higher accuracy. It is shown in Table (\ref{errorFEM}) that although the convergence rates of the two methods are almost the same at Curve 3 and Curve 5, mixed variation converges faster in general. These observations are in the general agreement with Babuska and Osborn\cite{babuska1978numerical} in their conclusion that the mixed quotient method should converge faster than the displacement based method when both methods use real basis. It would be interesting to know if and under what conditions, one basis leads to higher convergence over other basis. In our study, we found that (Table \ref{errorCexact} \& \ref{errorFEM}) trigonometric basis shows faster convergence on the 3rd and 5th branch, whereas the real basis converges faster on the 2nd branch.
\begin{table}[htp]
\caption{The average convergence rates, $\xi$, approximated using the results from $N=8$ to $28$ for the following four branches.}\label{errorFEM}
\centering
\begin{tabular}{lcccc}
\hline $\xi$ &  branch 2& branch 3& branch 4& branch 5\\
\hline Mixed quotient & 4.2845 & 1.1240 & 2.9316 &	0.9671\\
FEM &	3.0303 & 1.1066 & 2.0982 & 0.9447 \\
\hline
\end{tabular}
\end{table}
\section{Comparison with PWE}
One of the advantages of trigonometric expansion is that it can be made to immediately satisfy the Bloch boundary conditions. PWE is very similar to displacement Rayleigh quotient in the sense that both of them use the same momentum equation and trigonometric terms to express displacement, but, unlike Rayleigh quotient, PWE also expresses the material properties in trigonometric expansion. It is natural to consider which is more efficient. The material properties and reference solutions are the same as those in Section \ref{2D}. 
\begin{figure}[htp]
\centering
\includegraphics[scale=0.3]{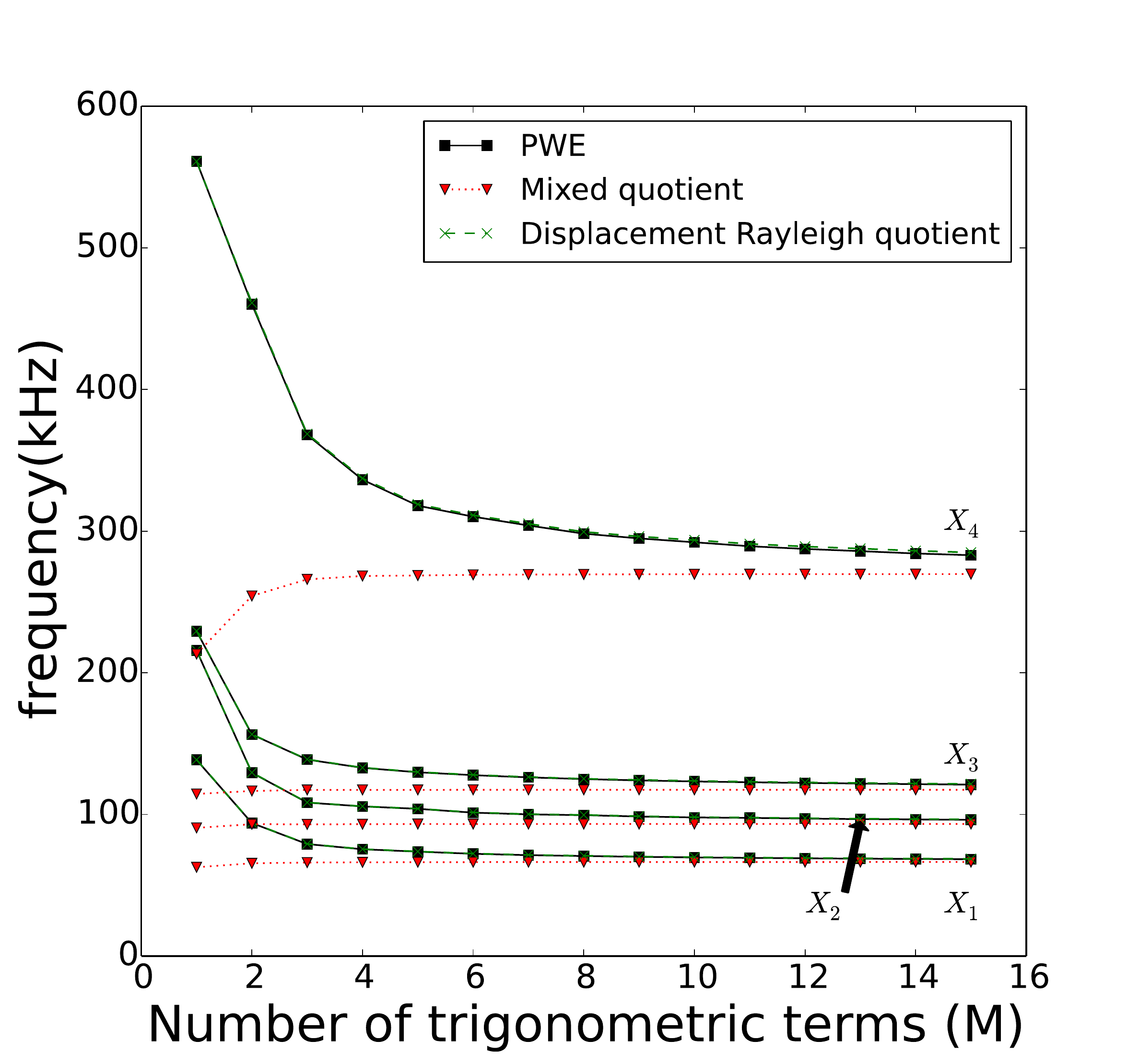}
\caption{Bandstructure calculated using Mixed quotient and PWE}\label{convergencePWE}
\end{figure}
PWE convergence rates are calculated for $(2\times M+1)^2$ plane waves, where $M=1$ to $15$ and the convergence results are provided in (Table (\ref{averageconvrPWE})). Comparing to the values in Table (\ref{averageconvr2D}), PWE converges slower to the reference solution than mixed quotient does, however, the convergence rates are very close to displacement Rayleigh quotient. This is also evident in Fig. (\ref{convergencePWE}), which shows the evolution of the first 4 solutions on the IBZ symmetric point $X$. It can be seen that PWE and displacement results are very close to  each other. Goffaux and S\'{a}nches-Dehesa \cite{goffaux2003two} conclude that Rayleigh quotient converges faster than PWE for higher branches. However, the differences in actual calculated values in their paper, as well as in ours, are minuscule. In fact, if convergence is calculated as in (\ref{ineqconvergentrate}), then PWE would be seen to converge slightly faster than Rayleigh quotient both in their paper and in ours (Table (\ref{averageconvrPWE})).
\begin{table}[htp]
\caption{This table includes the values of average convergence rates of PWE method, $\xi_P$, from curve 1 to curve 4 measured at $X$ and the midpoints of $X-\Gamma$, $\Gamma-J$, $J-X$. }\label{averageconvrPWE}
\centering
\begin{tabular}{l|cccc||l|cccc}
\hline $\xi_P$  & $X$ & $X-\Gamma$ & $\Gamma-J$ & $J-X$ &	$\xi_P$  & $X$ & $X-\Gamma$ & $\Gamma-J$ & $J-X$\\
\hline Curve 1 & 1.2948 & 1.2539 & 1.2651 & 1.2647 & Curve 2 & 1.2884 & 1.2209 & 1.1713 & 1.2690\\
\hline Curve 3 & 1.2074 & 1.2968 & 1.3046 & 1.2336 & Curve 4 & 1.1971 & 1.2325 & 1.2349 & 1.1817\\
\hline
\end{tabular}
\end{table}
\section{Conclusions}\label{conclusions}
We have presented comparative convergence studies of three variational principles used for solving elastodynamic eigenvalue problems. The formulations of three fundamental  principles presented in this paper can be easily applied to 1-, 2-, and 3-D periodic composites. The convergence behavior is seen to be related to the continuity and differentiability of the material property variations. Smoothly varying material properties in a unit cell result in higher convergence rates for all three quotients. Comparing the three methods, mixed quotient generally shows larger convergence rates for the different considered cases. Although generally showing superior effectiveness in convergence, mixed quotient is as fast as displacement or stress Rayleigh quotient when either compliance or density is constant respectively. The relative errors produced by the mixed quotient for a given $M$ can be smaller by as much as $M^{-2}$, when compared to those produced by the displacement or stress Rayleigh quotients using trigonometric terms. In the comparisons between mixed quotient and two other methods , FEM and PWE, we find that the mixed quotient, in general, has larger convergence rates. 

\section*{Acknowledgements}\label{acknowledgements}
This research has been conducted under UCSD/ONR W91CRB-10-1-0006 award to the Illinois Institute of Technology (DARPA AFOSR Grant RDECOM W91CRB-10–1-0006 to the University of California, San Diego).
\section*{References}


\end{document}